\documentclass[fleqn,10pt]{wlscirep}
\title{Structure constrained by metadata in networks of chess players}

\usepackage[normalem]{ulem} 
\usepackage{array,booktabs} 
\usepackage{bm} 

\author[1,2]{Nahuel Almeira}
\author[1,2]{Ana L. Schaigorodsky}
\author[2]{Juan I. Perotti}
\author[1,2,*]{Orlando V. Billoni}
\affil[1]{Facultad de Matem\'atica, Astronom\'ia, F\'isica y Computaci\'on, Universidad Nacional de C\'ordoba}
\affil[2]{Instituto de F\'isica Enrique Gaviola (IFEG-CONICET), Ciudad Universitaria, 5000 C\'ordoba, Argentina.}
\affil[*]{billoni@famaf.unc.edu.ar}


\begin{abstract}

Chess is an emblematic sport that stands out because of its age, popularity and complexity. It has served to study human behavior from the perspective of a wide number of disciplines, from cognitive skills such as memory and learning, to aspects like innovation and decision making.  Given that an extensive documentation of chess games played throughout history is available, it is possible to perform detailed and statistically significant studies about this sport.
Here we use one of the most extensive chess databases in the world to construct two networks of chess players. One of the networks includes games that were played over-the-board and the other contains games played on the Internet.
We study the main topological characteristics of the networks, such as degree distribution and correlations, transitivity and community structure. 
We complement the structural analysis by incorporating players' level of play as node metadata. Although both networks are topologically different, we show that in both cases players gather in communities according to their expertise and that an emergent rich-club structure, composed by the top-rated players, is also present.
\end{abstract}

\begin{document}

\flushbottom
\maketitle

\thispagestyle{empty}

\section*{Introduction}

Chess is an iconic ancient game with a development intimately related to the history of mankind and, in modern times, 
to the evolution of computers\cite{Prost12}. It is known that high levels of expertise require hard training even for 
the most talented players. Due to the popularity and complexity of chess~\cite{Ribeiro13PO,Atashpendar16EPL} this game 
has been used to study several  fields  of human behavior. For instance,  chess activity has been used to evaluate 
cognitive performance in professional and novice players \cite{Duan12PO}, decision making 
processes~\cite{Sigman10FN, Leone17C}, and the relation between expertise and knowledge~\cite{Chassy11PO,Sala2017}.
There are also interesting results regarding the collective behavior of a pool of players~\cite{Blasius09PRL,Perotti13EPL,Schaigorodsky14PA, Schaigorodsky16PO}. For example, by exploring chess databases Blasius and T\"{o}njes \cite{Blasius09PRL} observed that the pooled distribution of chess opening weights follows Zipf's law with a universal exponent, and explained these findings in terms of an analytical treatment of a multiplicative process. Moreover, studying the growing dynamics of the game-tree, we have found that the emerging Zipf and Heaps laws can be explained in  terms of nested Yule-Simon preferential growth processes~\cite{Perotti13EPL}. Also, we have observed interesting memory effects in the history of the recorded games~\cite{Schaigorodsky14PA, Schaigorodsky16PO}.

Nowadays there are big active communities of chess players that comprise all levels of expertise playing in both, the traditional over-the-board way and in online portals. These very large world-wide communities of chess players produce extensive game records, providing  a source of data useful for large-scale analyses. In these databases the expertise of each active player is statistically well characterized through the `Elo' system introduced by the physicist Arpad Elo~\cite{Glickman95ACJ}. Elo is not just a rating system that allows to rank the players according to their level of play but is also a predictive system that permits to anticipate their future performance. Hence is useful, for instance, for pairing players in tournaments and to follow their evolution.

Like in almost every human activity, it is expected that chess players gather together to form networks structured in communities. The statistical properties of this kind of networks, including their community structure, have not been yet characterized to the best of our knowledge. When considering chess players as nodes of a network, the expertise takes the role of node information, or \emph{metadata} and can give important information about the topological properties of the network. For instance, homophilic features are expected to be present in this kind of networks, as it is widely known~\cite{Acher2016} that players tend to play against similar skilled opponents.
Moreover, the relation between players' expertise and network structure can provide useful insights about the mechanisms that work in the formation of communities in social systems. 

The study of correlations between the topology of social networks and the metadata of its nodes is currently a prominent topic in complex networks~\cite{Leo16JRSI,Newman16NC,Hric2016,Hric14PRE}. Its importance arises because it can provide clues in the understanding of the structure and the dynamics of social communities. Moreover, the relation between network topology and metadata has been proved to be useful for detecting social stratification~\cite{Leo16JRSI}, disassortative communities~\cite{Newman16NC} and data inconsistency such as missing-links~\cite{Hric2016}. It has also served in pointing out possible drawbacks of the currently employed community detection algorithms~\cite{Hric14PRE}. However, the data available to study this types of correlations is yet scarce and there are just a few studies of social systems that are analyzed from this point of view.

The aim of this paper is to describe the community of chess players in a world-wide scale in terms of the theory of complex networks and to combine network topology with metadata in order to study the similarities and differences that exist between online and over-the-board playing. The results are divided in four sections. First, we introduce two chess databases from which we performed our analysis and provide some relevant statistical information about them. Second, we describe how the networks of chess players were constructed from the data. Third, we present a characterization of the networks, giving their main structural properties and discussing the relation between structure and associated node metadata. Last, we perform a community detection analysis and show the correlations between community membership and players' level of play. 

\section*{Results}

\subsection*{Statistics of the databases}
\label{sec:database}

\begin{figure}[ht]
\begin{center}
\includegraphics[scale=0.35]{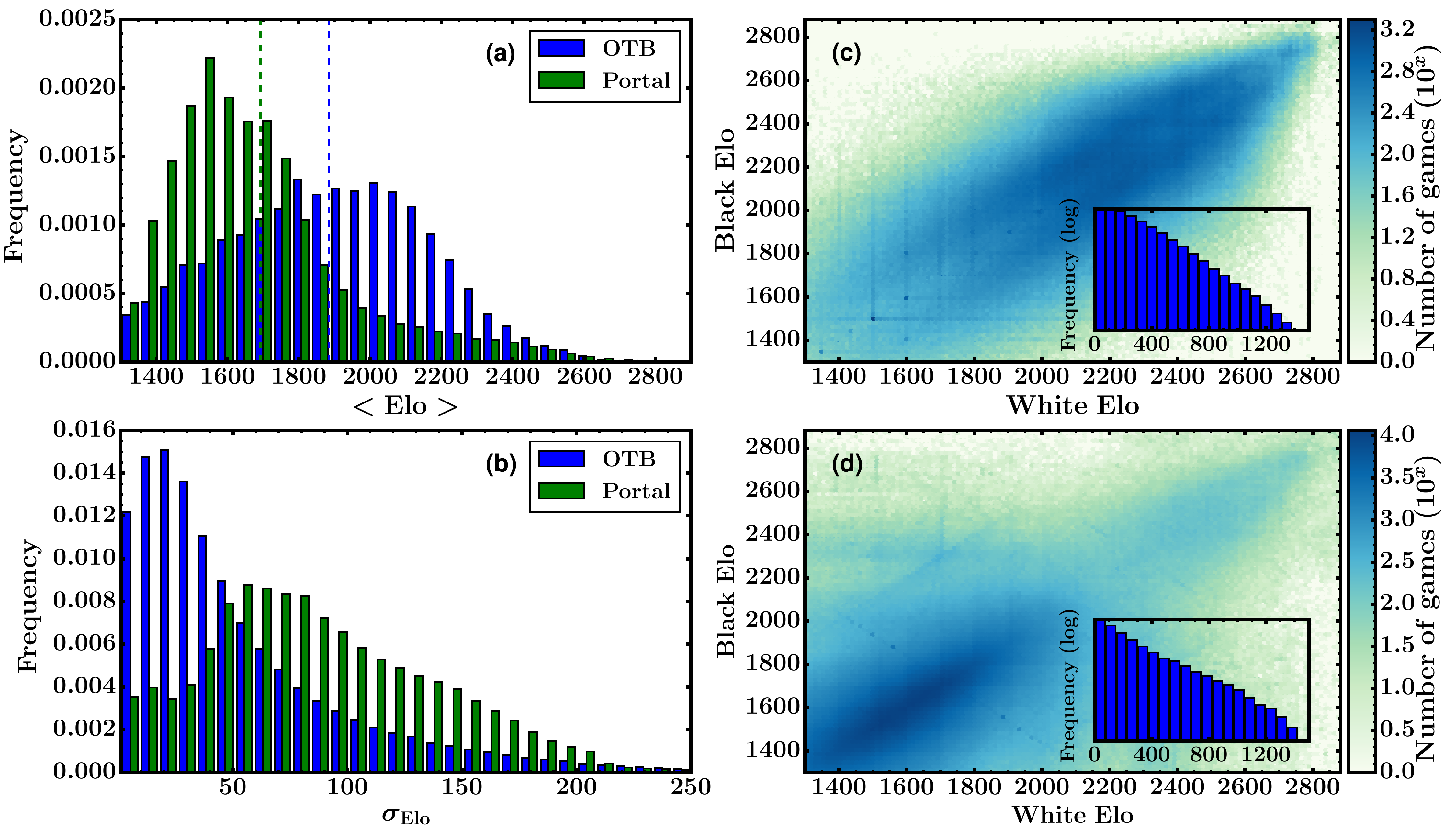}
\caption{\label{fig:elo_distribution} Elo statistics of over-the-board (OTB) and internet (Portal) databases --see Section \emph{Statistics of the databases for a description of each dataset.} \textbf{(a-b)} Mean and standard deviation distribution for players' Elo. Vertical dashed lines in \textbf{(a)} represent the mean value of each distribution. \textbf{(c-d)} Two dimensional density plot of the Elo values for White and Black players at the moment of a game. \textbf{(c)} OTB database, \textbf{(d)} Portal database. The histograms in the insets, presented in log-linear scale, indicate an exponential decay for the distribution of Elo differences.}
\end{center}
\end{figure}

Two databases were used to construct the networks. Both of them were provided by Opening Master, a chess database company that claims to have the world largest collection of chess games. The first database is a set of 7.7 million games played over-the-board, and we refer to it as `OTB'. The second one, which we call `Portal', 
contains more than $15$ million chess games played between humans in different websites or portals, such as chess.com, 
ICC-chessclub.com, Playchess.com, freechess.org and chesscube.com. Each database includes several information fields for 
each game, such as the name of the players, their Elo rating, the result, the date, the opening played and the sequence of moves.
Since the Elo of each particular player is a varying quantity (it is updated after each game), we computed each player's mean 
Elo value $\langle \mathrm{Elo} \rangle$ and its corresponding standard deviation $\sigma_{\mathrm{Elo}}$. 
In Figure \ref{fig:elo_distribution} \textbf{(a)} 
we show the mean Elo distributions for the complete set of players, computed for each database, and in Figure \ref{fig:elo_distribution} \textbf{(b)} we plot the corresponding distributions of the standard deviations. 
Both networks exhibit centered $\langle \mathrm{Elo} \rangle$ distributions that can be well approached by Gaussian distributions. The mean Elo value for OTB is $1884\pm276$ while for Portal it is $1692\pm228$.
Although we don't show it here, the time evolution of the Elo is interesting on its own. Some players, for instance older top-rated players, maintain a rather constant value, except for fluctuations. Others, for example outstanding young professional players, show a more or less monotonic increasing evolution. Despite it could bring to interesting results, a thorough study of the Elo dynamics is beyond the scope of this work.

In the game of chess, players prefer to face opponents that have a similar level of play, i.e., similar Elo~\cite{Acher2016}. One of the reasons for this is motivational; players in general are not interested in playing a game if their chances of winning (or losing) are excessively high. The other reason is related to the way the Elo score is adjusted after a game. If a player with high Elo wins a game against a lower rated opponent, she/he will gain just a few points of Elo, but if she/he loses the game, she/he will lose a large amount of points. As a consequence, high Elo players -specially the top ranked ones- try to avoid playing against lower ranked opponents. 
This behavior can be visualized from the data by employing a two dimensional density plot, corresponding to the values of Elo of White and Black players at the moment of a game, as depicted in Figure \ref{fig:elo_distribution} \textbf{(c-d)}. As it can be seen, most of the games where played between players whose Elo difference $\Delta E$ is not greater than 300. In fact, the Elo difference distribution is almost exponential, as can be seen in the insets of these figures. The mean value and standard deviation are $\langle \Delta E \rangle = 168,\, \sigma_{\Delta E} = 131$ for OTB and $\langle \Delta E \rangle = 145,\, \sigma_{\Delta E} = 156$ for Portal. As we move towards higher values of Elo, the dispersion narrows, meaning that top-rated players tend to form a closed community. A clear difference between both databases is that in OTB most of the games correspond to players with high Elo. On the contrary, in Portal most of the games are played between lower ranked players. 

It is worth to point out that the previous statement does not imply that in OTB most of the players are top-rated --actually, it can be seen from the histogram of figure \ref{fig:elo_distribution} \textbf{(a)} that this is not the case. Instead, this result only implies that the few very active top-rated players contribute more to recorded events than the rest of the players. This also explains why the average Elo in OTB is lower than the standard scores of professional players and close to Portal's mean value. In fact, in a recent published work~\cite{Schaigorodsky16PO} in which we analyze a different over-the-board database we also found that
a small fraction of players generate a significant fraction of the games recorded in the database.

\subsection*{Network construction} \label{sec:generation}

\begin{figure}[ht]
\begin{center}
\includegraphics[scale=0.3]{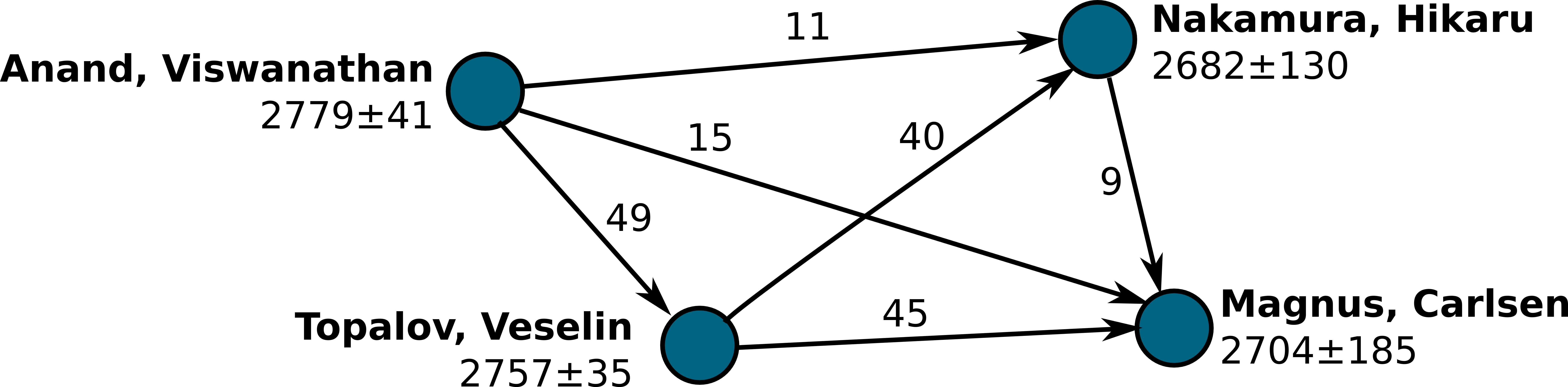}
\caption{\label{fig:example} Representation of the subnetwork composed by four top-rated players in the OTB database. Each node represents a player, containing as metadata the name of the player and her/his mean Elo (with the corresponding standard deviation). If two players have met al least once, they are joined by a link. The weight and the direction of the link depends on the net flux of Elo between those players, after counting all the games they played each other. Take, as an example,  players Anand and Carlsen. The existing link between them goes from Anand to Carlsen, and has a weight of $15$ meaning that, after considering all the games they have played against each other, the net result was that Anand transferred $15$ points of Elo to Carlsen.}
\end{center}
\end{figure}

With the information contained in the databases we built a weighted directed network for each database. As depicted in Figure \ref{fig:example}, each node in the network represents an individual player and contains two different metadata, namely the name of the player and her/his mean Elo. Links between nodes are established only if the players have played at least one game. Weights and directions of the connections are assigned according to the net flux of Elo between the players (each time a game is finished there is a net transfer of Elo from one player to the other, which depends on the result of the game and on the initial Elo of both players). For each pair of players we computed the net flux of Elo over all the games they have played against each other. If the net flux goes from player `A' to player `B' we added a link pointing from A to B with a weight equal to the flux value. We also computed a binarized version of the networks, i.e. the corresponding undirected and unweighed networks. Finally, we constructed different randomizations of the networks and employed them as null models (details on how the randomizations where performed can be seen in the Methods section).

\subsection*{Network Characterization}
\label{sec:characterization}

\begin{figure}[ht]
\begin{center}
\includegraphics[scale=0.35]{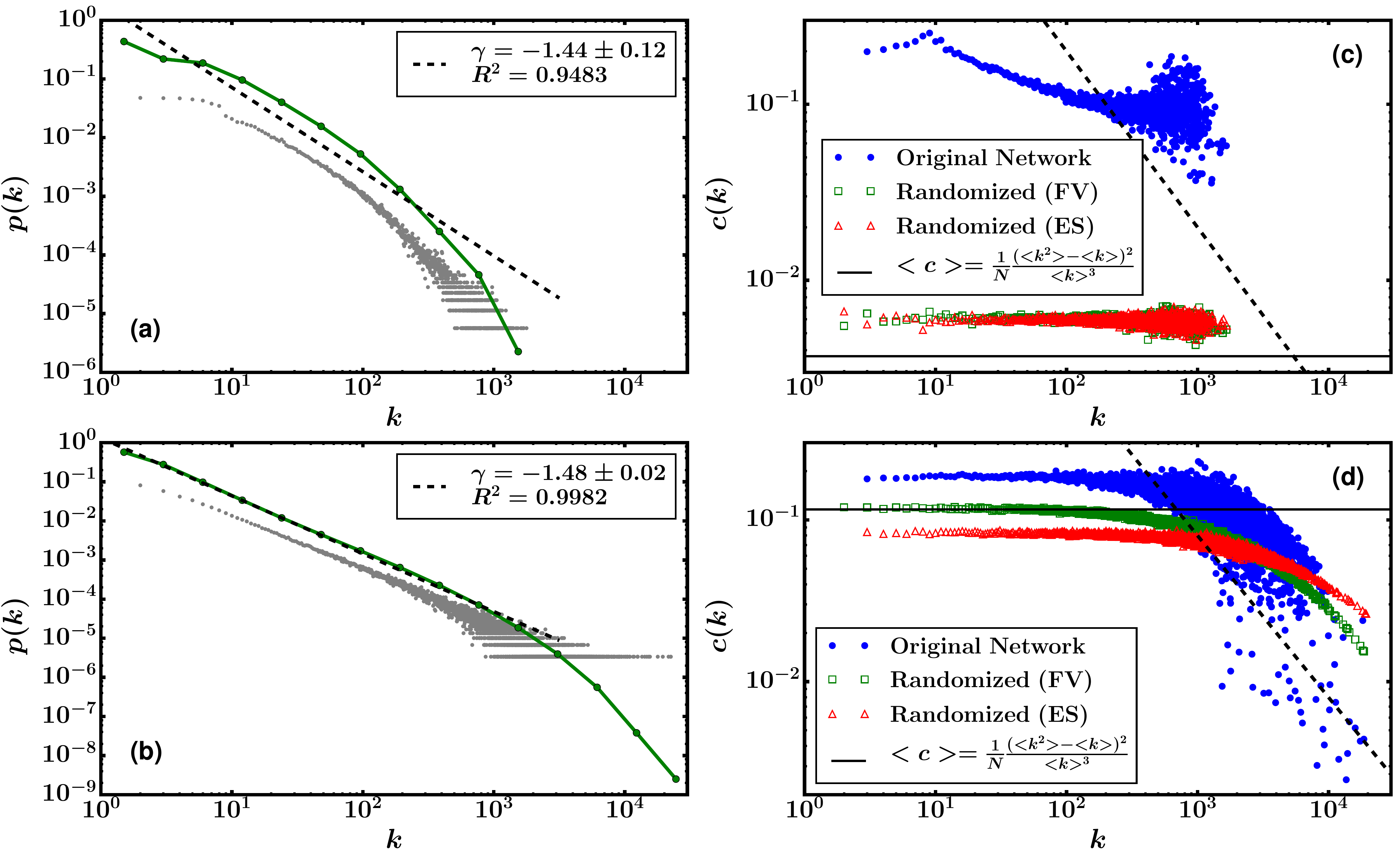}
\caption{\label{fig:degree_pdf} Structural properties of the networks. \textbf{(a-b)} Degree probability distribution function corresponding to \textbf{(a)} OTB and \textbf{(b)} Portal network. Grey points represent the linear bin histogram with bin size equal to 1 and green dots are the logarithmic histogram with scale factor equal to 2. The dashed black line corresponds to a linear fit of the logarithmic histogram. The slope $\gamma$ of the fit and its $R^2$ value are given in the legends. \textbf{(c-d)} Clustering coefficient as a function of the degree for \textbf{(c)} OTB and \textbf{(d)} Portal network. The figures also include the clustering coefficient computed for the randomized networks, using the algorithms of Fabien-Viger (FV) and double-edge swapping (ES). The horizontal line corresponds to the expected value for a fully uncorrelated network with the same degree distribution. The dashed line represent a dependence $c(k) \sim k^{-1}$ as expected for hierarchical networks.}
\end{center}
\end{figure}

We computed the degree distribution for the binarized version of both OTB and Portal networks. As shown in Figure \ref{fig:degree_pdf}, both probability distributions exhibit a right-skewed tail, feature that is shared among several social networks \cite{Barabasi1999,Newman2001}. Although similar, they exhibit some differences. While the OTB distribution is concave in the log-log plot, the degree distribution of the Portal network is well fitted by a power law with an exponent of $\sim 1.5$ across 3 orders of magnitude, followed by a gentle decay at the tail. As the original networks are directed and weighted, we computed the node strength~\cite{Yook2001} distribution. The results (not shown) are similar to the ones obtained for the degree distribution.

In order to explore the presence of degree correlations we computed the Pearson correlation coefficient $r$ as introduced by Newman\cite{Newman2002} (see Methods). We observe (Table \ref{meassures_table}) that OTB is assortative, whereas Portal exhibits a slightly disassortative behavior. As it has been previously mentioned in the literature~\cite{Newman2002}, many social networks with a long tailed degree distribution are assortative like, for example, the science coauthorship and the film actor collaboration networks~\cite{Newman2003}. The noteworthy fact here is that Portal network strays from this behavior and behaves instead more like a technological or a biological network~\cite{Newman2002}.

\begin{table}
  \centering
  \renewcommand{\arraystretch}{1.5}
  \begin{tabular}{>{\centering\bfseries}m{0.5in} |>{\centering}m{0.6in} >{\centering}m{0.7in} >{\centering\arraybackslash}m{0.4in} >{\centering\arraybackslash}m{0.4in} >{\centering\arraybackslash}m{0.4in} >{\centering\arraybackslash}m{0.4in} >{\centering\arraybackslash}m{0.4in} |>{\centering\arraybackslash}m{0.4in} >{\centering\arraybackslash}m{0.4in} >{\centering\arraybackslash}m{0.4in}}
    \toprule
    Network & $N$ & $M$ & $\langle k \rangle$ & $\langle k^2 \rangle/\langle k \rangle$ & $C$ & $C_{WS}$ & $r$ & $Q$ & $\mu$ & $N_C$\\
    \midrule
    OTB & $184,566$ & $3,256,880$ & $34$ & $194$ & $0.100$ $(0.006)$ & $0.181$ $ (0.006)$ & $0.396$ $(0.001)$ & $0.69$ $(0.16)$ & $0.24$ $(0.88)$ & $3,604$ $(1,416)$\\
    Portal & $296,660$ & $8,529,638$ & $58$ & $1,408$ & $0.085$ $(0.061)$ & $0.178$ $(0.083)$ & $-0.106$ $(0.000)$ & $0.42$ $(0.10)$ & $0.40$ $(0.93)$ &  $12,734$ $(3,861)$\\
    \bottomrule
  \end{tabular}
  \caption{Different statistical measures computed for each studied network. $N$: number of nodes, $M$: number of
         links, $\langle k \rangle$: mean degree, $\langle k^2 \rangle/\langle k \rangle$: heterogeneity parameter, $C$: clustering coefficient (Newman) and $C_{WS}$: clustering coefficient (Watts-Strogatz), $r$: assortativity parameter, $Q$: modularity, $\mu$: mixing parameter and $N_C$: number of communities. The values of $Q$, $\mu$ and $N_C$ were obtained by employing the Louvain algorithm. Values between parentheses correspond to randomized networks using the double-edge swapping method.}
        \label{meassures_table}
\end{table}

We studied the transitivity within the networks\cite{NewmanBook} by computing their clustering coefficients, following the definitions of Newman\cite{NewmanBook} and Watts-Strogatz~\cite{Watts1998} (see Methods). As can be seen in Table \ref{meassures_table}, both networks have similar values for these measures. The differences between OTB and Portal arise when comparing them with their random networks, as randomization substantially decreases the transitivity for OTB but has almost no effect on Portal.
We also computed the clustering coefficient $c(k)$ as a function of the degree~\cite{Vazquez02PRE}. In Figure \ref{fig:degree_pdf}, we show that for both networks $c(k)$ is a decreasing function, and that decreases more notoriously for Portal than for OTB.
The behavior, which has been previously observed in cases such as world-trade~\cite{Fagiolo07PRE} and metabolic networks \cite{Ravasz2002S},  has been reported as a good indicator of an existing hierarchical structure~\cite{Ravasz2003,Ravasz2002S} in the network. In particular, the original model proposed by Ravasz, predicts a power law dependence $c(k)\sim k^{-1}$ for a hierarchical network~\cite{RavaszThesis}. The dotted lines in Figures \ref{fig:degree_pdf} \textbf{(c-d)} indicate this relation

Since in disassortative networks the clustering of hubs is bounded from above \cite{serrano2005tuning}, 
disassortative networks are less clusterized than assortative ones\cite{foster2011clustering}. This is consistent with our results (OTB is assortative and more clusterized than Portal, which is disassortative), supporting a more pronounced hierarchical structure in Portal than in OTB.
After randomization, correlations for $c(k)$ completely disappear in OTB, whichever randomization algorithm is employed. In Portal, the two randomization algorithms we employed substantially diminish dispersion and reduce correlations, but none of them can fully remove them.
%

\begin{figure}[ht]
    \centering
        \includegraphics[scale=0.27]{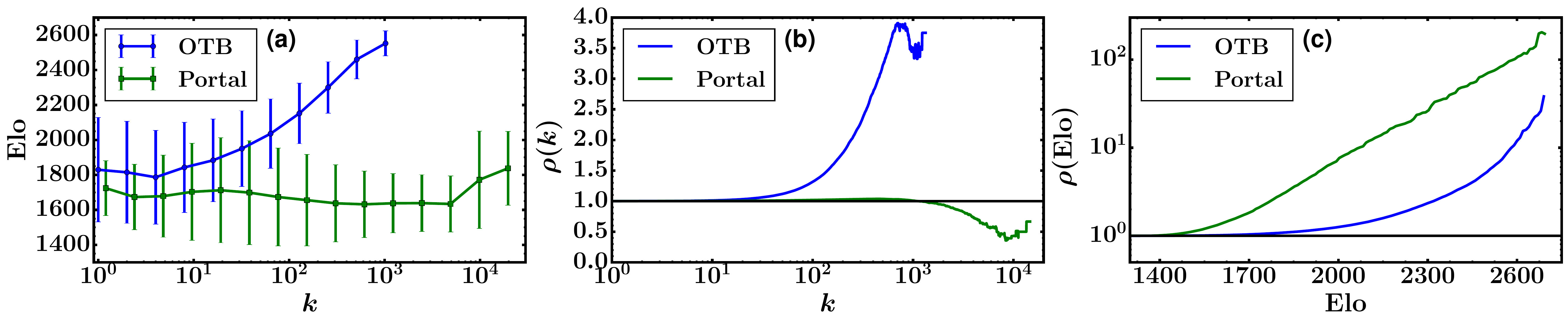}
        \caption{Interplay between network structure and node metadata. \textbf{(a)} Correlation between the players' mean Elo and their degree for both networks. A positive correlation is seen for OTB but in Portal the two variables seem not to be correlated. \textbf{(b)} Normalized rich-club coefficient defined as a function of the degree. A notorious positive rich-club phenomenon is seen for OTB but Portal barely strays from the null model. \textbf{(c)} Normalized rich-club coefficient defined as a function of the Elo. Both networks show a significant rich-club phenomenon when considering this node property. For comparison, a straight line for $\rho = 1$ (the rich-club coefficient for a fully randomized network) is plotted in the last two figures.}
        \label{fig:rich_club}
\end{figure}

Another aspect of the networks that we studied was the relation between nodes degree and the mean Elo of the corresponding players.
According to Figure \ref{fig:rich_club} \textbf{(a)}, in OTB the number of opponents each player meets is positively correlated to the average rank of the player and negatively correlated to its variability. On the other hand, no such correlations are observed in Portal.
We also analyzed the normalized rich-club coefficient~\cite{Zhou2004,Colizza06NL} $\rho(k)$ defined in terms of the degree (see Methods) for both networks. Results are shown in Figure \ref{fig:rich_club} \textbf{(b)}, where it can be observed that the degree-based rich-club phenomenon is present in the OTB network whilst is absent in Portal. Moreover, higher degree nodes in Portal are less connected to each other as compared to random networks. 
As we discuss in the Methods section, the rich-club coefficient can be also defined in terms of a quantitative node metadata, such as the mean Elo of the players. Figure \ref{fig:rich_club} \textbf{(c)} shows that, under this definition, a rich-club structure emerges in both networks. This fact indicates that there exists a densely connected group, composed by the best players, in both chess communities. The individuals that belongs to this group, nevertheless, are not necessarily those who meet more opponents. 

\subsection*{Communities}                      
\label{sec:communities}

\subsubsection*{Communities based on modularity}

In many complex networks the distribution of links tends to be inhomogeneous at different levels of organization. At a microscopic level, inhomogeneities manifest themselves in the form on long-tailed degree distributions. At a mesoscopic level, certain groups of nodes are found to be densely connected to each other and relatively separated from the rest of the network. Such groups of nodes are said to form \emph{communities} --also called \emph{clusters} or \emph{modules}--, which can give useful information about the network, because it is probable that nodes that belong to the same community share common properties or play similar roles in the system\cite{Hric2016}.

Given a network and a particular partition, i.e. a function that assigns each node to a given community, one can compute the so called \emph{modularity} \cite{Clauset2004} $Q$. This property aims to measure how good the clustering is by evaluating the number of links 
that connect nodes lying within the same communities. If the fraction of inner links does not differ 
from what would be expected in a random network, the value of the modularity is zero. According to Clauset et. al., a value 
over 0.3 indicates that the network has a significant community structure \cite{Clauset2004}. Another important quantity that measures whether the communities in a network are well defined or not is the mixing parameter \cite{Fortunato2010} $\mu$, which is the fraction of inter-community links. More specifically, small (large) values of the mixing parameter indicate that the communities are well (loosely) defined. 

In order to study the community structure of our networks we employed the Louvain \cite{Blondel2008} algorithm, which can be used with weighted networks. The choice of the method was based on both its performance and its speed (for a comparison among several state-of-the-art algorithms, see~\cite{Yang2016}). The community structures obtained were characterized by means of their modularity, mixing parameter, number of communities and community size distribution. The first three measures are summarized in Table \ref{meassures_table}. We observed that communities are better defined in the OTB network as compared to Portal. This can be seen by noting that OTB has a larger value for $Q$ and a lower value for $\mu$ than Portal. As expected, randomization considerable increases the mixing parameter and decreases the modularity in both networks. The number of communities is also reduced in the randomized networks.

\begin{figure}[ht]
    \centering
	\includegraphics[scale=0.35]{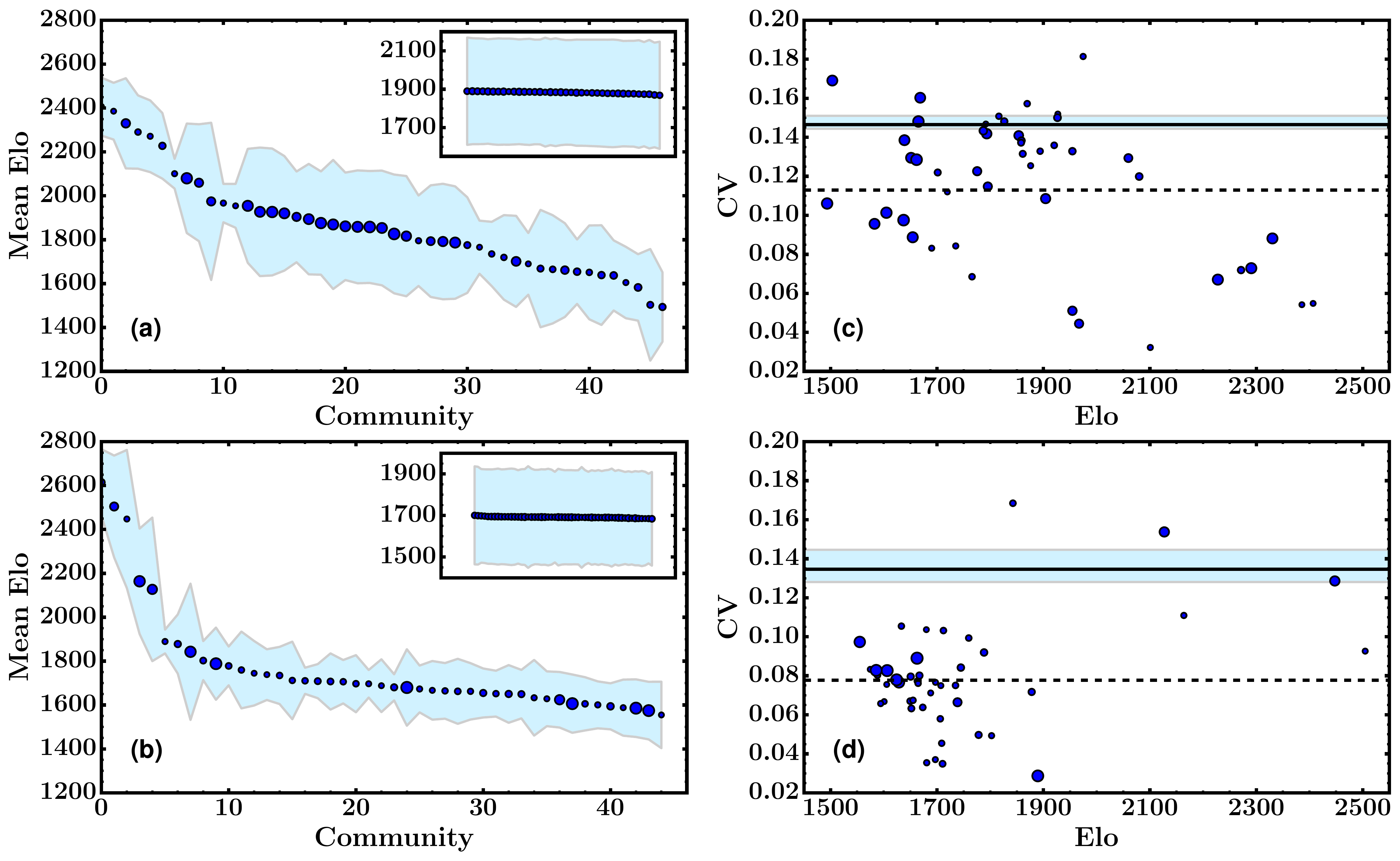}
	\caption{Correlations between mean Elo and community membership. Each dot represents a community, with a marker size proportional to the community size. \textbf{(a-b)} Communities ranked according to the mean Elo (from higher to lower values). The shaded region represents the standard deviation of the Elo of each community. Insets show the results obtained for the respective randomized networks, using the double-edge swapping method. \textbf{(c-d)} Coefficient of variation versus mean Elo of the communities. Horizontal solid lines correspond to the coefficient of variation computed for the whole networks. The filled regions represent the CV range in which every community detected in the corresponding random network is contained. The dotted line correspond to the center of mass of the CV by weighting each community by its size. Top plots belong to the OTB and bottom plots to Portal.}
        \label{fig:meanEloVsComm}
\end{figure}

We computed the mean Elo of the pool of players that belong to each community, and then ranked the communities according to 
that value. We observed that the range of Elo spanned by the communities is approximately contained in the interval $(1400,\;2600)$. As shown in Figure \ref{fig:meanEloVsComm} \textbf{(a-b)}, two regimes can be identified for each network in this ranking plot. Each of them is characterized by a linear relation between ranking and Elo, but with a different slope. For top-rated communities, the mean Elo steeply decreases as the ranking decreases, indicating a strong stratification of the communities. For lower ranked communities the slope shallows. As can be seen in the figure, this effect is more notorious in Portal than in OTB. After randomization, the effect completely disappears. As shown in the insets of the figures, the mean Elo is nearly the same for all communities and the standard deviation becomes larger than in the original networks.

Correlations between Elo and communities can be better visualized by computing the coefficient of variation CV of the Elo, that is, the quotient between the standard deviation of the Elo and its mean. In Figure \ref{fig:meanEloVsComm} \textbf{(c-d)} we plot the CV for each community as a function of its mean Elo. It can be observed that the variability of the Elo is, on average, lower inside the communities than it is for the whole network. Moreover, by comparing the results to those obtained for the null models (see filled region in the figures), it can be seen that the difference is indeed significant. To put it in other words, Louvain algorithm is able to successively discriminate players according to their expertise. 

\subsubsection*{Communities based on metadata}

In order to explicitly detect correlations between communities and the Elo of the players we analyzed the community structure using an algorithm recently introduced by Newman and Clauset~\cite{Newman16NC} (see Methods). The algorithm employs Bayesian inference to construct a generative model that links community structure of the network with node metadata. As opposed to the Louvain algorithm, the number of communities is in this case fixed to a pre-defined value $K$. We used, as before, the mean Elo as the metadata. The correlation between Elo and community membership was measured via the normalized mutual information NMI (refer to the Methods section).

\begin{figure}[ht]
    \centering
\includegraphics[scale=0.3]{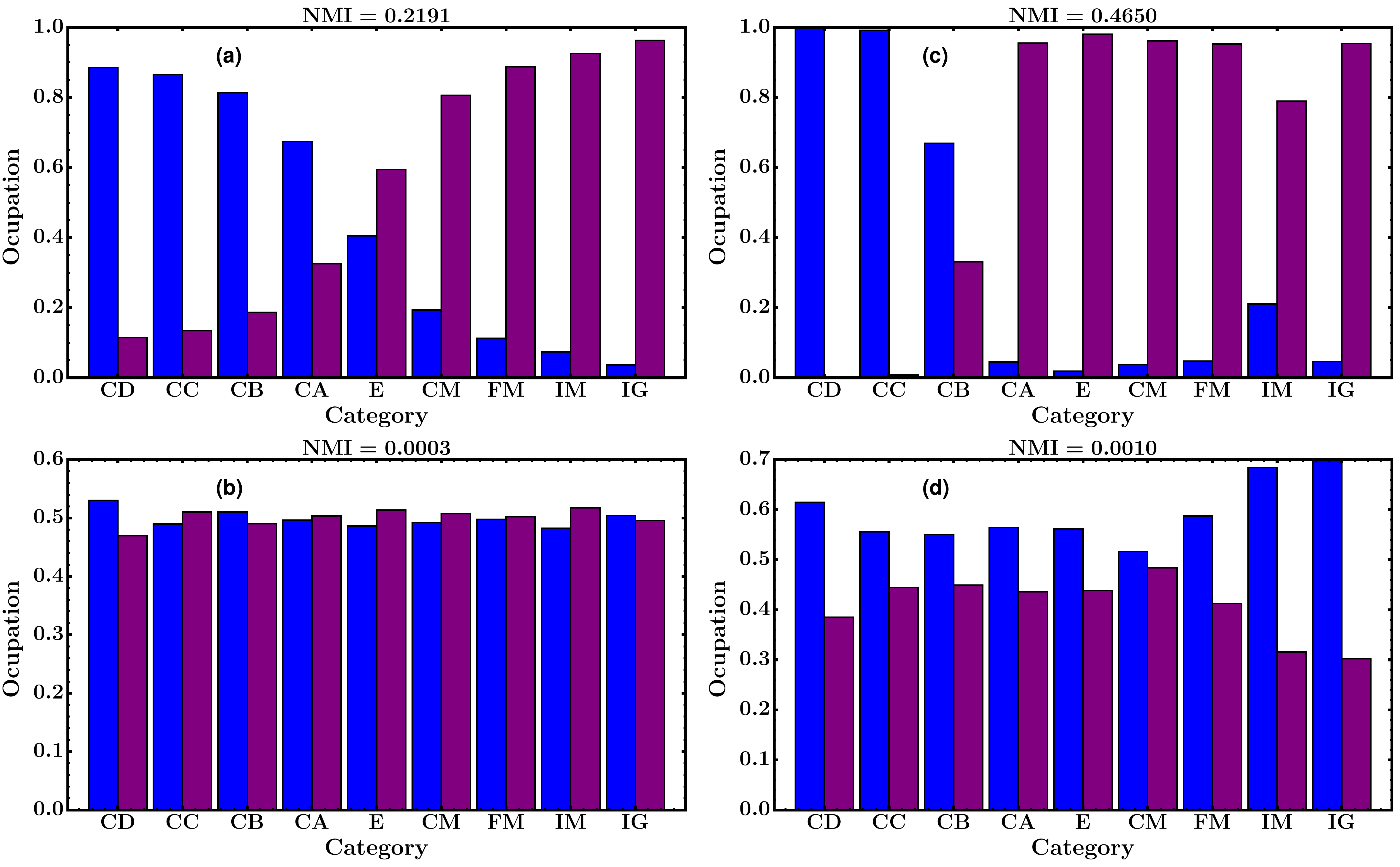}
	\caption{Distribution of Elo of each community detected by the algorithm of Newman and Clauset with $K = 2$. The color identifies the communities and the letters of the `Category' axis represent categories defined by a given range of Elo, from lower to higher values (see the  Methods section for a detailed definition of the categories).  Figure \textbf{(a)} corresponds to OTB and figure \textbf{(b)} to Portal. Figures \textbf{(c)} and \textbf{(d)} are the distributions for corresponding networks after randomizing the Elo of the players. Corresponding values for the NMI are given at each graph's title. }
    \label{fig:metadata}
\end{figure}

When the number of communities is fixed to 2, the Elo distribution in each community is clearly different; one community includes mostly low Elo players and the other high Elo players, as can be seen in Figure \ref{fig:metadata}. This feature is observed in both OTB and Portal networks, being most notorious in the second one. Now, it could be argued that the communities found are just an artifact of the algorithm employed. In order to test this hypothesis we repeated the experiments using three different network randomizations or null models. First, we performed, as before, a double-edge swapping. With this approach, the algorithm becomes particularly sensitive to the random seed, finding spurious communities that may or may not be correlated with the Elo. Second, we shuffled node metadata, i.e. we randomly reassigned the Elo of the players. This case is depicted in Figures \ref{fig:metadata} \textbf{(c-d)}, where it can be seen that the correlation between Elo and  communities disappears. Third, we performed a double randomization; after swapping edges we randomized the Elo. This case gives results similar to those obtained in the first case. We repeated the analysis by varying the number of communities from 3 to 9 (not shown) and obtained results that are qualitatively similar to the ones obtained for $K = 2$.

\section*{Discussion}

In addition to the traditional over-the-board (OTB), Internet portals have introduced a new way of playing chess, creating in the process two parallel communities of players. Although the game is the same, both communities evolved by their own, presenting important similarities and differences. When analyzing OTB games, it can be seen that recorded games are focused on professional players and include almost no amateur gaming. This is expected to happen because most of the amateur activity takes place in informal contexts where games are not recorded. As a consequence, players in OTB have large average Elos that slowly vary in time. On the Internet, this aspect is completely different, since every game is recorded --even if it has been carried on by players with low Elo. Thus, players on the Internet exhibit comparatively lower Elo values and greater variances.

When considering chess communities as complex networks, where nodes represent players and where links represent transfer of Elo between them, it can be seen that they have long-tailed degree distributions, meaning that there exist a huge variation between the number of games (and thus, the number of opponents) each player plays in her/his career. In the case of Internet games (Portal), the distribution is well fitted by a power law with exponent $\sim 1.5$, and a gently decay at the tail. The decay for 
large values of $k$ can be understood by the fact that chess players cannot play with an arbitrarily large number of opponents throughout their life, but they are constrained by the duration of their career.
In the case of OTB, the curve strays from a line in a log-log plot, presenting a concavity. 
The plateau for lower values of $k$ can be understood in terms of the discussion of the previous paragraph. Most of the players who play very few games are amateurs, so their games are not always recorded and thus, they do not appear in the databases.


One of the main differences between the studied networks is related to their degree-degree correlations. Whilst OTB is an assortative network --feature that seems to be common to the majority of social networks~\cite{Newman2002}--, Portal exhibits a dissasortative structure.
%
This difference can be understood by introducing players' level of play (i. e. their Elo) to the networks as node \emph{metadata} and considering the relationship between this measure and the number of opponents each player faces (the degrees of the corresponding nodes). In OTB, there is a positive correlation between these two variables, meaning that the most active players also turn out to be the best ones. Given that players face opponents with similar Elo, assortativity in terms of the degree becomes natural. On the other hand, Portal exhibits no correlation between Elo and degree, so it is equally likely for a player to face opponents with any level of activity.
A complementary approach was performed by studying the rich-club phenomenon for each network. We observed that in both chess communities (OTB and Portal), elite players tend to form a densely connected group, which is reflected on a normalized rich-club coefficient that increases as a function of the Elo. 

In terms of network transitivity, we showed that both networks present a decreasing value of the clustering coefficient as a function of the degree. Nevertheless, whilst in OTB this decrease is rather slow, it becomes more steep in Portal, being well approached by a relation $c(k) \sim k^{-1}$. As discussed by Ravasz~\cite{Ravasz2003,Ravasz2002S,RavaszThesis}, this suggests a more pronounced hierarchical structure in Portal. This is consistent with the fact that correlations between clustering and degree cannot be fully removed by performing a network randomization, indicating that the observed transitivity is inherent to the topology of Portal. 
{The presence of this hierarchical structure implies that hubs do not form a closed community. This is consisting with the result obtained from the rich-club analysis in terms of the degree, which shows that the normalized rich-club coefficient decays for higher degrees.
The isolation of hubs is possible because they belong to the class of intermediate level players, which is the most popular one. In chess portals most of the games involve two players picked at random among the pool of players with similar Elo. Since there are few hubs, it is unlikely that in a random pairing they face each other forming closed communities.


The existence of communities in OTB and Portal networks was confronted against corresponding experiments performed with appropriate null models, which consisted in randomized versions of the networks preserving the degree distributions of the original ones. From this analysis we observed that the modularities of the null models are considerably lower than those of the corresponding original networks, and that the opposite relation occurs for the mixing parameter. 
The community detection analysis also confirmed the existence of a correlation between the community structure of the networks and the Elo of the players. As a general rule, the Elo of a community is less dispersed than the Elo of the complete network. Supporting the previous statement, the results obtained by employing the algorithm proposed in~\cite{Newman16NC} show that the Elo is significantly correlated to the community structure of the networks.

\section*{Methods}

\subsubsection*{Network randomization}

In order to compare our results with a null model, we performed two different randomization of the networks. 
In both cases we rearranged the links in a way that maintains the degree distribution invariant. One of the 
randomization employed was the Fabien-Viger~\cite{Viger2015} algorithm, which is based on the known 
Molloy-Reed~\cite{Molloy1995,Molloy1998} random graph generating model. It allows to construct a simple and connected 
random graph with an arbitrary degree distribution. Each realization of the algorithm gives a sample of the 
ensemble of such graphs, taken with uniform probability.
The second randomization was based on double-edge swapping. The procedure is as follows. Suppose node $u$ is 
connected to node $v$ and node $w$ is connected $x$. If $u$ is not connected to $w$ and neither $v$ is 
connected to $x$, edges $uv$ and $wx$ are removed and replaced by edges $uw$ and $vx$. This step is repeated 
as many times as needed so as to fully randomize the network. Based on Milo, et. al. criterion~\cite{Milo2003}, we performed 
a number of steps equal to ten times the number of links of each network.

\subsubsection*{Clustering coefficient and correlations}

In social networks, it is common to see that when a node $u$ is connected to another node $v$, and $v$ is connected to a third node $w$, 
it is likely that $u$ is also connected to $w$. In common parlance, this means that ``the friend of my friend is also my friend''. 
This characteristic is called \emph{transitivity} or \emph{clustering} and can be quantified in different ways~\cite{Watts1998,NewmanBook}. 
In this work, we employed the so called clustering coefficient and  computed two definitions that are widespread in the literature. According to Watts and Strogatz~\cite{Watts1998}, the  
clustering $C_{WS}(i)$ of node $i$ can be defined as:
\begin{equation}
 C_{WS}(i) = \dfrac{e_i}{k_i(k_i-1)/2},
\end{equation}
where $k_i$ is the degree of node $i$ and $e_i$ is the number of edges between his neighbors. The average clustering coefficient of the complete network is then
\begin{equation}
 C_{WS} = \dfrac{1}{N} \sum^{N}_{i = 1} C_{WS}(i).
\end{equation}
Alternatively, the clustering coefficient of the complete network can be defined as\cite{NewmanBook}:
\begin{equation}
 C = \dfrac{3 \times \text{(number of triangles)} }{\text{number of connected triples}},
\end{equation}
where connected triples means three nodes connected by at least two edges. Finally, as pointed out in~\cite{Vazquez02PRE}, it is sometimes worthy to discriminate the clusterization of the network according to  
the degree of the nodes. Thus a degree dependent clustering coefficient can be defined as 
\begin{equation}
c(k) = \dfrac{1}{N_{k}} \sum^{N_k}_{i/k_i = k} C_{WS}(i),
\end{equation}
where the sum ranges over all nodes in the network having degree $k$ and $N_k$ is the number of such nodes.

Degree-degree correlations can be calculated through the Pearson coefficient\cite{Newman2002}. This coefficient ranges from $-1$ and $1$, and allows to quantify if nodes with similar degree are more prone to be connected to each other than nodes with very different degrees. In the first case, $r>0$ and the network is said to be assortative. In the latter case, nodes are connected if its degrees are very different, hence $r<0$ and the network is dissasortative. 

\subsubsection*{Rich-club phenomenon}

The rich-club phenomenon is related to the tendency of nodes of high degree to connect to each other forming tightly 
interconnected communities. As proposed by Zhou and Mondrag\'on~\cite{Zhou2004}, this effect can be measured by introducing 
a coefficient defined as follows:
\begin{equation}
 \phi(k) = \dfrac{2 M_{>k}}{N_{>k} (N_{>k}-1)},
\end{equation}
where $M_{>k}$ and $N_{>k}$ are the remaining links and nodes in the network after removing nodes with degree smaller or equal to a given value $k$. As it was mentioned by Colizza~\cite{Colizza06NL}, this coefficient is a monotonically increasing function even for uncorrelated networks, so it is necessary to compare it with the one obtained from an appropriated null model 
in order to detect a truly rich-club tendency. To correction of this spurious effect is obtained by defining $\rho(k) = \phi(k)/\phi_{ran}(k)$, where $\phi_{ran}$ is the rich-club coefficient of a fully randomized network. In the context of social systems a normalized rich-club coefficient $\rho(k)$ that increases with degree $k$ corresponds to the existence of a kind of oligarchy in the organization form of the system.  

The rich-club can be also defined in terms of a node property different from the degree. In this paper, we choose the mean Elo for each player and defined the rich-club phenomenon for the Elo property as
\begin{equation}
 \phi(\mathrm{Elo}) = \dfrac{2 M_{\mathrm{>Elo}}}{N_{\mathrm{>Elo}} (N_{\mathrm{>Elo}}-1)},
\end{equation}
where $M_{\mathrm{>Elo}}$ and $N_{\mathrm{>Elo}}$ are the remaining links and nodes in the network after removing nodes with mean Elo smaller or equal to a given value $\mathrm{Elo}$. From this definition we derived the normalized coefficient $\rho(\mathrm{Elo})$ in the same way as done before.

\subsubsection*{Communities based on metadata}

Newman and Clauset~\cite{Newman16NC} introduced a community detection algorithm that makes use of Bayesian statistical inference to construct a generative network model possessing the specific features one hopes 
to find in the data, namely community structure and a correlation between that structure and node metadata. 
Then the model is fit to an observed network having metadata and the parameters of the fit give the information about the structure of the network.

The model is based on the stochastic block model~\cite{Holland1983}, but incorporates the dependence on node metadata via a set of prior probabilities. An important fact is that the number of communities to be find is a fixed parameter, in contrast with, for example, the Louvain algorithm. It is also worthy to stress that this algorithm does not suppose any \emph{a priori} relation between structure and metadata, but if there exist any, the algorithm should be able to find it. Among the different alternatives of quantifying correlation between communities and metadata we employed here, as proposed by the authors, the normalized mutual information NMI between the membership of nodes to each metadata category and their membership to each community found by the algorithm. In order to incorporate the metadata to the algorithm, the mean Elo was divided into a set of categories, inspired in the nomenclature used by the Word Chess Federation. The corresponding 
assignation rule of range of Elo and category is summarized in Table \ref{tab:categories}.

\begin{table}
  \centering
    \begin{tabular}{c|c|c}
    Category  & Abbreviation  & Range of Elo  \\
    \hline
    \hline
    International Grandmaster & IG  & $>2500$  \\
    International Master & IM     & $2400-2500$  \\
    FIDE Master & FM & $2300-2400$ \\
    Candidate Master & CM  & $2200-2300$ \\
    Expert & E  & $2000-2200$  \\
    Class A & CA & $1800-2000$  \\
    Class B & CB & $1600-1800$  \\
    Class C & CC & $1400-1600$  \\
    Class D & CD & $1200-1400$  \\
    \end{tabular}
  \caption{Assignation rule for the metadata categories based on the Elo.}
  \label{tab:categories}
\end{table}

\subsection*{Data availability}

The datasets that support the findings of this study are available from Opening Master (www.openingmaster.com) but restrictions apply to the availability of these data, which were used under license for the current study, and so are not publicly available. Data are however available from the authors upon reasonable request and with permission of Opening Master.

\section*{Acknowledgements}

We thank Opening Master for providing the datasets. This work was partially supported by grants from CONICET (PIP 112 20150 10028), SeCyT--Universidad Nacional de C\'ordoba (Argentina).

\section*{Author contributions statement}

N.A performed the data analysis. All authors discussed the results. N.A and O.B. wrote the manuscript, with input from all the authors. 



\section*{Competing financial interests} The authors declare that they have no competing interest. 

\end{document}